\documentclass[prd,amsmath,amssymb,floatfix,notitlepage]{revtex4}
\usepackage{graphicx}
\usepackage{hyperref}
\usepackage{bm}
\usepackage{amssymb}
\begin{document}

\title{Information Content of Spontaneous Symmetry Breaking}

\author{Marcelo Gleiser}
\email{mgleiser@dartmouth.edu}
\affiliation{Department of Physics and Astronomy, Dartmouth College,
Hanover, NH 03755, USA}

\author{Nikitas Stamatopoulos}
\email{nstamato@dartmouth.edu}
\affiliation{Department of Physics and Astronomy, Dartmouth College,
Hanover, NH 03755, USA}

\date{\today}

\begin{abstract}
We propose a measure of order in the context of nonequilibrium field theory and argue that this measure, which we call relative configurational entropy (RCE), may be used to quantify the emergence of coherent low-entropy configurations, such as time-dependent or time-independent topological and nontopological spatially-extended structures. As an illustration, we investigate the nonequilibrium dynamics of spontaneous symmetry-breaking in three spatial dimensions. In particular, we focus on a model where a real scalar field, prepared initially in a symmetric thermal state, is quenched to a broken-symmetric state. For a certain range of initial temperatures, spatially-localized, long-lived structures known as oscillons emerge in synchrony and remain until the field reaches equilibrium again. We show that the RCE correlates with the number-density of oscillons, thus offering a quantitative measure of the emergence of nonperturbative spatiotemporal patterns that can be generalized to a variety of physical systems.

\end{abstract}

\maketitle

\section{Introduction}

Spatially-localized solutions to partial differential equations \cite{Scott} play a central role in many current models of particle physics \cite{Rajamaran} and condensed-matter physics \cite{Nelson}. Some of the most studied of these solutions are topological and nontopological defects that may appear during spontaneous symmetry breaking \cite{Peskin}. As is well-known, topological defects owe their stability to the nontrivial topology of the vacuum manifold \cite{Vilenkin}, while nontopological defects owe theirs to the conservation of a global charge \cite{Lee}. In both cases, solutions involve one or more interacting fields and are either time-independent or have a simple harmonic time-dependence $\sim \exp[i\omega t]$, as is the case of $Q$-balls \cite{Coleman} and other nontopological solutions \cite{Friedberg}.

Another class of localized-energy solutions, known as oscillons, exhibits nontrivial time dependence \cite{Bogolyubsky,Gleiser1}. During the past decade or so \cite{Oscillons}, oscillons were shown to exist in models with a single self-interacting scalar field in dimensions $d\leq 6$ \cite{Gleiser2} and in many models with gauge fields, Abelian \cite{GleiserThor} and nonAbelian \cite{GrahamMIT,GleiserThor2}, including the Standard Model \cite{GrahamSM}. They can be exceedingly long-lived, self-supported by their nontrivial interactions due to feedback from parametric resonance \cite{GleiserHowell,GleiserThor2}. Oscillons can also play an important role in cosmology, as reported in recent work \cite{GGS1,GGS2,Amin}. 

Both topological and nontopological defects and oscillons can be thought of as attractors in field configuration space \cite{GleiserSic}: given that certain dynamical constraints are satisfied, for a broad range of initial conditions the system will evolve toward these solutions. This is not necessarily surprising, since these configurations are solutions to the equations of motion. Semi-classically, we expect them to dominate the path integral. 

We may, however, ask whether there is another way to quantify the existence of nontrivial spatially-localized solutions in field configuration space. Can we think of spatially-coherent field configurations as ordered states in an informational sense, in analogy with the Shannon entropy of information theory \cite{Shannon,Brillouin}? That is, given the set of field modes that are allowed by the constraints of the model (i.e. initial and boundary conditions), do they carry a special informational signature that can be quantified?

To this end, we recently proposed a measure of configurational entropy, based on the Fourier transform $\phi(k)$ of square-integrable, bound functions $\phi(x)$ \cite{GleiserStamEnt}. Leaving the details for Section III, here it suffices to say that with this configurational entropy, we can establish a correlation between the energy of a localized-field configuration and its associated configurational entropy. In particular, we can show that departures from the solution of the eom will have correspondingly larger relative configurational entropies. 

In the present work, we take this approach one step further, applying it to nonequilibrium fields. Our recent treatment of Ref. \cite{GleiserStamEnt} was for static solutions to the field equations, such as $1d$ kinks and $3d$ bounces. Here, we will add time dependence, computing the relative configurational entropy during spontaneous symmetry breaking in the context of a $3d$ scalar field model. ``Relative'' refers to a comparison between the entropy of the field at some time and the entropy of the initial state, which we take to be a thermal state at temperature $T$. We will compute the change in relative configurational entropy as the field is tossed out of equilibrium during the symmetry-breaking process. Previous work has shown how, for certain types of quench, oscillons naturally emerge during symmetry breaking \cite{GleiserHowell,GleiserThor2}. Our results show quantitatively that the emergence of these localized coherent structures coincide with the largest departures from equilibrium and that they carry the most information content, in the sense defined in Section III. We are thus proposing a measure of order in field configuration space. Furthermore, we are able to establish a direct correlation between the emergence of ordered spatiotemporal structures, in effect the number density of such structures, and the relative configurational entropy.

The work is organized as follows. In Section II we describe the model and its lattice implementation. In Section III we define the relative configurational entropy (RCE) and apply it to spontaneous symmetry breaking. We describe how the RCE provides a measure of ordering in the system as compared to the maximally-disordered initial thermal state and how this order correlates with the existence of spatially-coherent structures. In Section IV we  present our conclusions and plans for future work. In the Appendix, we provide some technical details of the lattice implementation.

\section{The Model}
We consider a (3+1)-dimensional scalar field theory with Lagrangian density

\begin{equation}
 \mathcal{L}=\frac{1}{2}(\partial_\mu\phi)^2-V(\phi),
\end{equation}
and the tree level potential given by

\begin{equation}
 V(\phi)=\frac{m^2}{2}\phi^2-\frac{\alpha}{3}\phi^3+\frac{\lambda}{8}\phi^4,
\end{equation}
where the parameters $m$, $\alpha$, $\lambda$ are positive-definite and temperature independent. We use $\hbar=c=k_B=1$ and rescaled variables $\phi'=\phi\sqrt{\lambda}/m$, $x'^{\mu}=x^{\mu}m$, $\alpha'=\alpha/(m\sqrt{\lambda})$ to write the potential as $V(\phi)=(m^4/\lambda)V(\phi')$, with

\begin{equation}
 V(\phi')=\frac{\phi'^2}{2}-\alpha'\frac{\phi'^3}{3}+\frac{\phi'^4}{8}.
\label{rescaledPotential}
\end{equation}
We henceforth drop the primes and work with the rescaled variables. In this work we only consider the values $\alpha=0$ and $\alpha=3/2$. The first case corresponds to a potential with a single minimum at $\phi=0$ and the second describes a double-well potential with degenerate minima at $\phi=0$ and $\phi=2$. For $\alpha=0$ the potential is ${\cal Z}_2$-symmetric. For any other value of $\alpha$ this symmetry is broken. In the context of $2d$ \cite{GleiserHowell} and $3d$ \cite{GleiserThor2} models, it has been shown that when quenching a thermalized field from the symmetric to the double-well potential (here, from $\alpha=0$ to $\alpha= 3/2$), large-amplitude fluctuations about the vacuum state give rise to oscillon formation as the system evolves towards thermal equilibrium in the new potential. In other words, coherent, spatially-extended configurations develop spontaneously when an initially featureless system is tossed out of equilibrium. The mechanism behind this process is well understood: coherent oscillations of the field's zero mode parametrically amplify higher $k$-modes and the resulting energy transfer triggers the formation of oscillons \cite{GleiserHowell,GleiserThor2}. This remains true when the expansion of the universe is incorporated into the dynamics \cite{GGS1,GGS2}. We note that we could have investigated the traditional symmetry-breaking mechanism with a double-well symmetric about $\phi=0$. In this case, the ${\cal Z}_2$ symmetry-breaking would lead to spinodal decomposition and the formation of domain walls. 

\subsection{Lattice Implementation}

We simulate the formation of oscillons using a cubic lattice with $N^3=256^3$ points, periodic boundary conditions, lattice spacing $dx=0.5$ and time step $dt=0.01$. We prepare the initial thermal state using standard Langevin dynamics \cite{langevin}

\begin{equation}
 \ddot{\phi}+\gamma\dot{\phi}-\nabla^2\phi=-\frac{\partial V}{\partial \phi}+\zeta,
\end{equation}
where $V(\phi)$ is the rescaled potential in Eq.~\ref{rescaledPotential} with $\alpha=0$ and $\zeta$ is a Markovian noise with two-point correlation function obeying the fluctuation-dissipation relation

\begin{equation}
 \langle\zeta(\mathbf{x},t)\zeta(\mathbf{x}',t')\rangle=2\gamma T\delta(\mathbf{x}-\mathbf{x}')\delta(t-t'),
\end{equation}
where $T$ is the temperature parameter characterizing the initial state and we take $\gamma=1$. The system was evolved until thermalization was achieved, indicated by the onset of equipartition with every mode having average kinetic energy $T/2$, after which the potential was quenched from $\alpha=0$ to $\alpha=3/2$ and the field's coupling to the heat bath was removed ($\gamma\rightarrow 0$). So, after the quench the dynamics is conservative. The evolution of the field was done using a symplectic velocity Verlet algorithm and the Laplacian was discretized with a second-order accurate, forth-order isotropic stencil using all 26 neighbors of a $3\times 3 \times 3$ cube around a point \cite{isotropic}. We have checked that our results showed no particular dependence on the choice of box size, lattice spacing and timestep.

\subsection{Hartree Approximation and Oscillon Emergence}
Thermal fluctuations in $\phi$ will change the potential $V(\phi)$ which, to leading order in perturbation theory, can be approximated by the Homogeneous Hartree Approximation \cite{hartree}. Since the Hartree approximation assumes that the fluctuations of the field remain Gaussian throughout its evolution, it works well just before and after the quench for all temperatures. For low temperatures, it remains valid at all times. We can thus derive the Hartree potential by writing the field as $\phi=\bar{\phi}+\delta\phi$ and then averaging over all fluctuations $\delta\phi$ to get $V_H=\langle V(\bar{\phi}+\delta\phi)\rangle$. Under the Hartree assumptions we have $\langle\delta\phi\rangle=0$ and $\langle\delta\phi^2\rangle=\beta$, where $\langle\delta\phi^2\rangle$ is the mean square variance of the field and is proportional to the temperature parameter $T$. Suppressing the bar, the Hartree potential becomes

\begin{equation}
 V_H(\phi)=-\alpha\beta\phi+\left(1+\frac{3}{2}\beta\right)\frac{\phi^2}{2}-\alpha\frac{\phi^3}{3}+\frac{\phi^4}{8}.
\end{equation}
When thermal equilibrium is reached through the coupling to the heat bath, the field modes in momentum space satisfy

\begin{equation}
 \langle|\phi_{\rm eq}(k)|^2\rangle=\frac{T}{k^2+m_H^2},
\label{equilibrium}
\end{equation}
 with the Hartree mass given by $m_H^2=V''_H(0)=\left(1+\frac{3}{2}\beta\right)$.

On the lattice, Eq.~\ref{equilibrium} has to be adjusted for lattice effects due to the $\phi$ dependence on the discretization and the inherent UV cutoff on the lattice. For the lattice UV cutoff $\pi/dx$, $\beta$ can be analytically obtained in terms of the temperature $T$ as \cite{GleiserThor2}  

\begin{equation}
 \beta=\frac{3T}{4\pi dx}.
\end{equation}
The continuous dispersion relation $\omega^2=k^2+m_H^2$ has to be modified to take into account the field dependence on the discretization scheme. For the isotropic discretization we use here, the radially-averaged dispersion relation will be given by $\omega^2=k_{\textrm{eff}}^2+m_H^2$, with \cite{lattice-dispersion} 

\begin{eqnarray}
 k_{\textrm{eff}}^2&=&-\frac{c_4}{dx^2}-\frac{2}{\pi dx^2}\int_0^{\pi/2}\cos\theta\int_{0}^{\pi/2}\big[ 2c_3\left[\cos \left[k\cos \phi \cos \theta dx\right]+\cos [k\sin \phi \cos \theta dx]+\cos [k\sin \theta dx]\right]\nonumber \\
&& +4c_2\left[\cos \left[k\cos \phi \cos \theta dx\right]\cos [k\sin \phi \cos \theta dx]+\cos \left[k\cos \phi \cos \theta dx\right]\cos [k\sin \theta dx]+\cos [k\sin \phi \cos \theta dx]\cos [k\sin \theta dx]\right]\nonumber\\
&& +8c_1\cos \left[k\cos \phi \cos \theta dx\right]\cos [k\sin \phi \cos \theta dx] \cos [k\sin \theta dx]\big]d\phi d\theta,
\label{k_eff}
\end{eqnarray}
and $c_1=1/30$, $c_2=1/10$, $c_3=7/15$, $c_4=-64/15$ being the discretization coefficients of the Laplacian. With these lattice effects taken into account, the two-point correlation function at equilibrium becomes

\begin{equation}
 \langle|\phi_{\rm eq}^{\rm latt}(k)|^2\rangle=\frac{T}{k_{\textrm{eff}}^2+1+9T/(8\pi dx)}.
\label{equilibrium-lattice}
\end{equation}

\begin{figure}[htbp]
\includegraphics[width=0.49\linewidth]{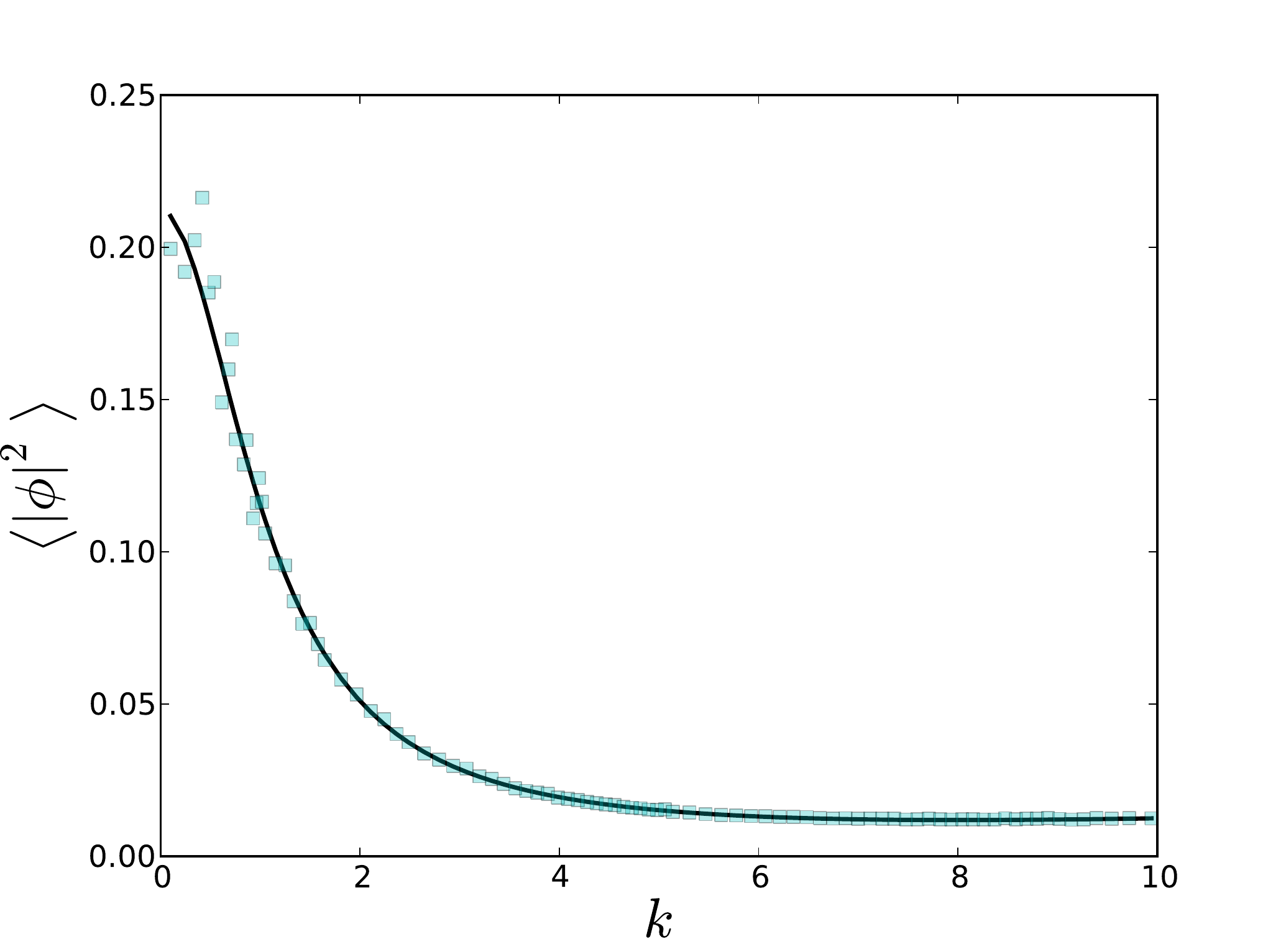}
\includegraphics[width=0.49\linewidth]{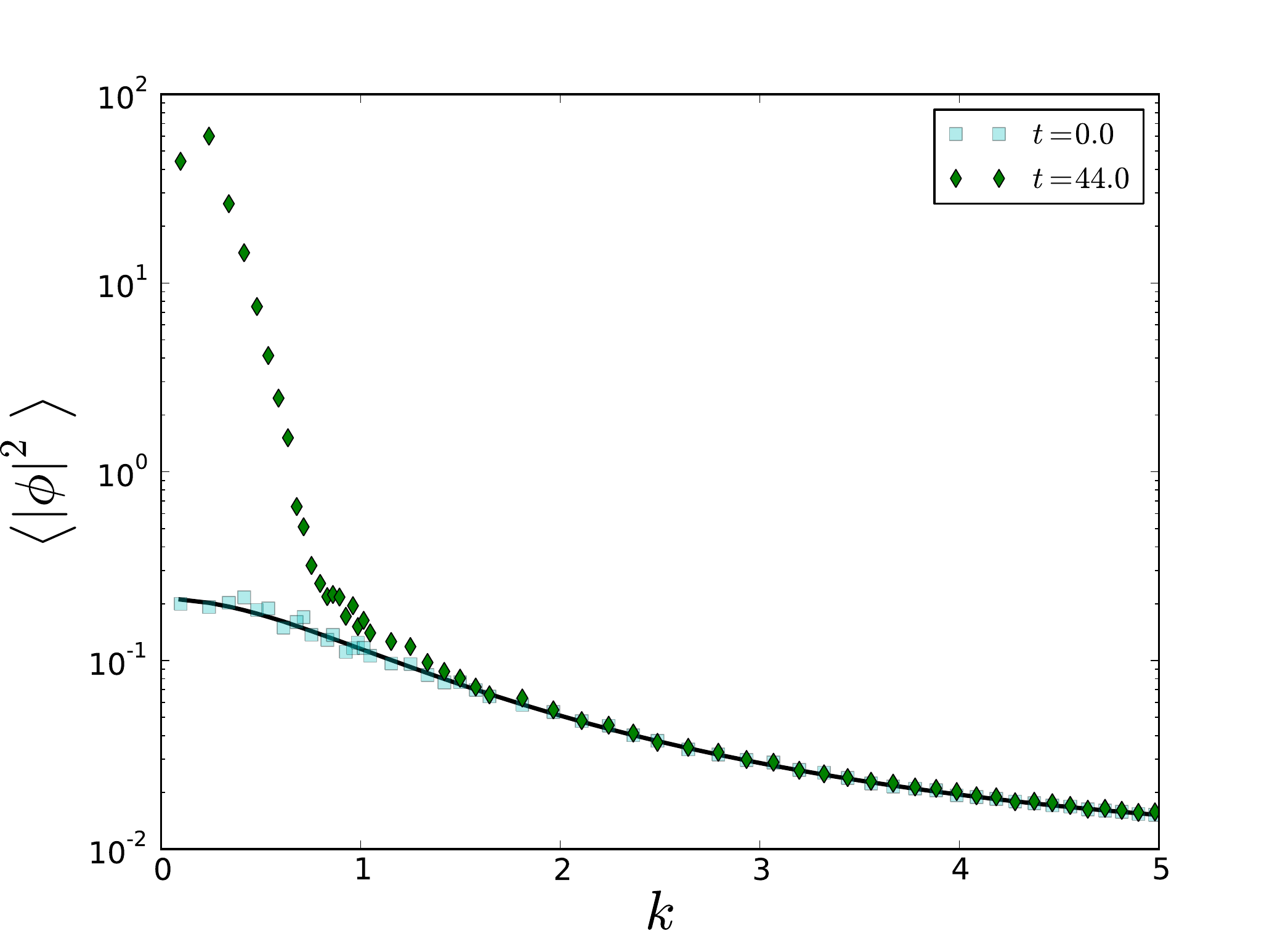}
\caption{Radially-averaged two-point correlation functions for the field $\phi$ at $T=0.25$. On the left, we show the simulation data at equilibrium and the analytical approximation $\langle|\phi_{\rm eq}^{\rm latt}(k)|^2\rangle$ (solid line) of Eq.~\ref{equilibrium-lattice}. On the right, the same data is plotted along with the two-point correlation function data at time $44m^{-1}$ after the quench (diamonds). Modes with $k\lesssim 1.5m$ get significantly amplified while the rest remain in thermal equilibrium throughout the simulation.}
\label{Fig:spectrum}
\end{figure}

\begin{figure}[htbp]
\includegraphics[scale=0.5]{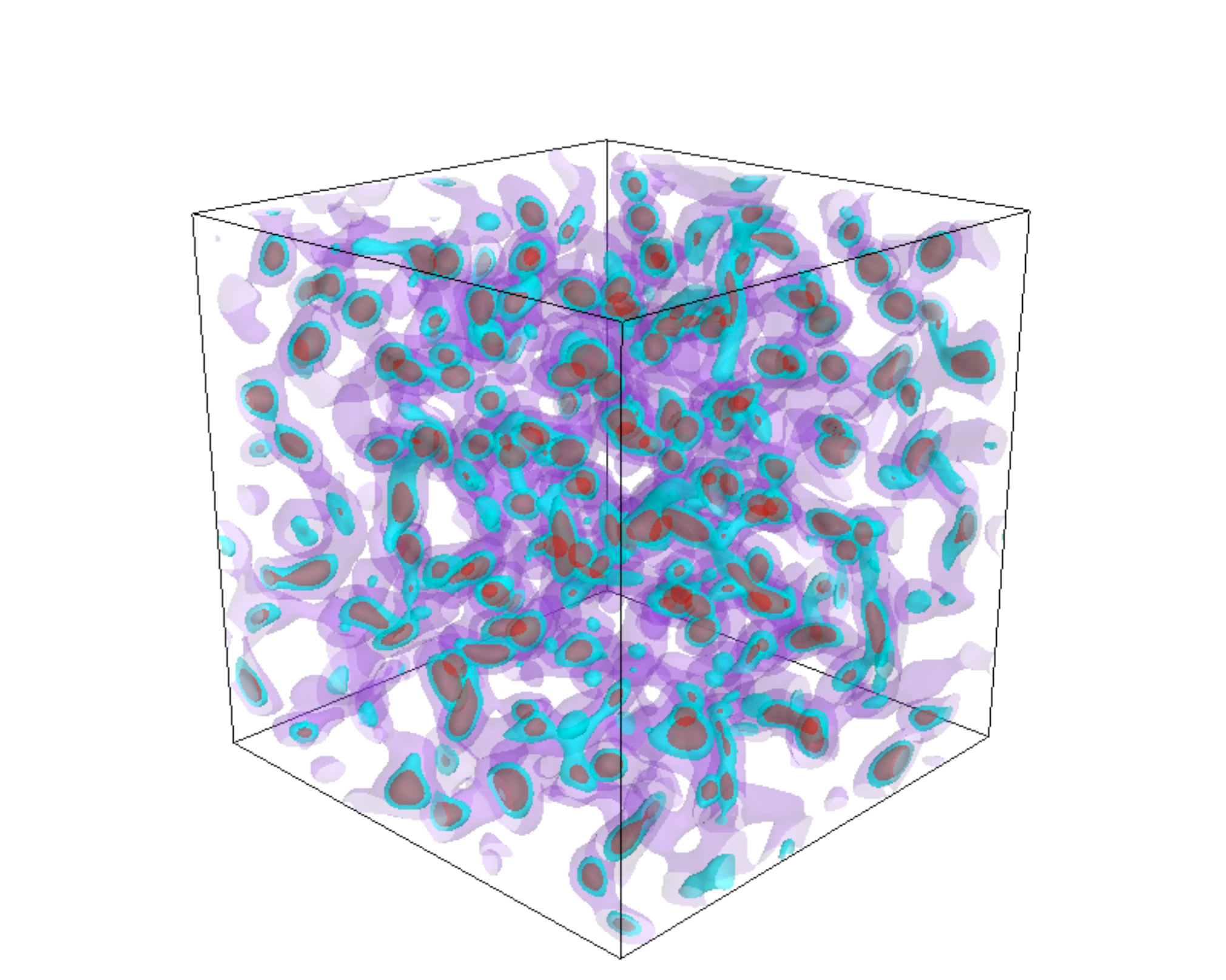}
\caption{Snapshot of the field $\phi$ at $t=44m^{-1}$ after the system is quenched from a single to a symmetric double-well. The simulation size shown here is $L^3=128^3$. The field was initially thermalized at $T=0.25$. Oscillons (spatial ordering) appear as spikes about the zero mode of the field. Three isosurfaces are shown at $\phi={0.5,1.3,1.6}$ in purple, cyan and red, respectively. As the system evolves towards its new equilibrium, the spatiotemporal ordering is gradually lost and oscillons subsequently disappear. A full visualization of the process can be seen at \cite{simulation}, where it is also clear that for early times oscillons emerge in synchrony (time ordering).}
\label{Fig:oscillons}
\end{figure}

In the left part of Fig.~\ref{Fig:spectrum} we show the radially-averaged two-point correlation function for the field $\phi$ after it reaches thermal equilibrium at temperature $T=0.25$. Squares denote the data from the numerical simulation, and the black solid line is the theoretically predicted spectrum $\langle|\phi_{\rm eq}^{\rm latt}(k)|^2\rangle$, adjusted for lattice effects. The averaging for low $k$-modes is less accurate because, on the lattice, there are fewer modes to compute the average of the power spectrum. As $k$ gets larger, there are more modes with the same value of $k$ and the agreement with the theoretically-predicted spectrum is evident. On the right we plot the equilibrium spectrum (squares) and the spectrum at $t=44m^{-1}$ (diamonds) after the quench. Modes with $k\lesssim 1.5m$ have been parametrically amplified, whereas modes with $k\gtrsim 1.5m$ have remained in thermal equilibrium. The low $k$, out of equilibrium modes that are parametrically amplified after the quench are the ones responsible for oscillon formation. Oscillons emerge synchronously after the quench as the system is tossed out of equilibrium, and then slowly disappear as the field evolves towards its new equilibrium state \cite{GleiserHowell, GleiserThor2}. A visualization of the process can be seen at \cite{simulation}. In Fig.~\ref{Fig:oscillons}, we show a snapshot of the field at $t=44m^{-1}$ after a thermal Wiener filter has been applied to remove the high $k$-modes still in equilibrium, with oscillons appearing as localized spikes in the field.

\section{Information Content of Coherent Field Configurations}
Reverting back to Fig.~\ref{Fig:spectrum}, we observe a marked difference in the two-point correlation function of field modes between the equilibrium (left) and nonequilibrium phases (right). In particular, as we emphasized before, low $k$-modes with $k\lesssim 1.5 m$ are greatly amplified when the field is away from equilibrium \cite{GleiserThor2}. In this section, we propose a measure to quantify this nonequilibrium amplification of low $k$-modes. We further show that this measure, having a natural interpretation as an entropy in field configuration space, correlates with the number of coherent field configurations (oscillons, in the case studied here) that emerge as the system is tossed out of equilibrium. Thus, we propose that our entropic measure can be used to quantify the emergence of complexity in field theory, if by complexity we understand the appearance of spatially-localized coherent field configurations, in contrast with the structureless (disorganized) thermal state. 

Following our recent work \cite{GleiserStamEnt}, we define the modal fraction of a field $\phi({\bf x},t)$ in Fourier space at time $t$ as

\begin{equation}
 f({\bf k},t)=\frac{|\phi({\bf k},t)|^2}{\int |\phi({\bf k},t)|^2 d^3k}.
\label{equilibrium_modal}
\end{equation}
Equivalently, in equilibrium we have

\begin{equation}
 g({\bf k})=\frac{|\phi_{\rm eq}({\bf k})|^2}{\int |\phi_{\rm eq}({\bf k})|^2 d^3k},
\end{equation}
with $|\phi_{\rm eq}({\bf k})|^2$ given by Eq.~\ref{equilibrium}. Note that with this definition, $g({\bf k})$ depends on the temperature $T$ only through the Hartree mass correction term $\beta$. In the lattice implementation, $g({\bf k})$ is computed with $|\phi_{\rm eq}^{\rm latt}({\bf k})|^2$ as defined in 
Eq. ~\ref {equilibrium-lattice}.
The linear dependence on the temperature is cancelled by our choice of normalization, which gives the modal fraction units of $m^{-3}$. Next, we define the dimensionless relative configurational entropy (RCE) of the power spectra as

\begin{equation}
 S_f(t) = \int S_f({\bf k},t)d^3k,
\label{relative_entropy}
\end{equation}
where the relative configurational entropy density $S_f({\bf k},t)$ is given by 

\begin{equation}
S_f({\bf k},t)=f({\bf k},t)\ln\frac{f({\bf k},t)}{g({\bf k})}.
\end{equation}
Our definition of the RCE is a field-theory version of the Kullback-Leibler divergence (KLd) commonly used in information theory to compare two probability distributions $P$ and $Q$ of a discrete random variable: $D_{\rm KL} = \sum P_i\ln (P_i/Q_i)$ \cite{kullback-leibler}. In information theory, the KLd  gives a measure of the expected number of extra bits required to code samples from $P$ using a code based on $Q$. Usually, $P$ represents the ``true'' data or a precisely computed distribution, while $Q$ represents a theory, model, or approximation of $P$.

By comparing the modal fraction of the field's Fourier transform at time $t$ with that of the thermal state, we obtain a measure of the amplification of the low $k$-modes responsible for oscillon formation, a ``distance'' in Fourier configuration space from the thermal state. (In principle, other mode expansions could be used, although, as we argued in Ref. \cite{GleiserStamEnt}, for localized fields the Fourier transform is the most natural.) Modes that remain in equilibrium throughout the evolution of the field (here, roughly for $|k|>1.5m$) have $f(k,t)=g(k)$ and $S_f(k,t)=0$, and thus do not contribute to the RCE. The larger the modal fraction $f(k,t)$, the larger its associated relative entropy density $S_f(k,t)$. 

The RCE thus provides a clear measure of the departure from equilibrium. Furthermore, it peaks where the field is most organized into coherent spatial structures.  This is consistent with the notion that the farther a system is from equilibrium, the farther it is from satisfying equipartition. In effect, since the thermal state has maximum entropy and hence no information (all modes have the same average energy--as equipartition determines) the RCE gives a measure of information in field configuration space: peaks in the RCE correspond to peaks in information-rich coherent structures. In the case here of a scalar field with an attractive self-interaction, the attractor point of its dynamics, the oscillon, is the farthest that it can be from equipartition. This explains why the RCE peaks when oscillons are present, with the amplitude of the peaks correlating directly with the number of oscillons, as we will see below.

\begin{figure}[htbp]
\includegraphics[width=0.49\linewidth]{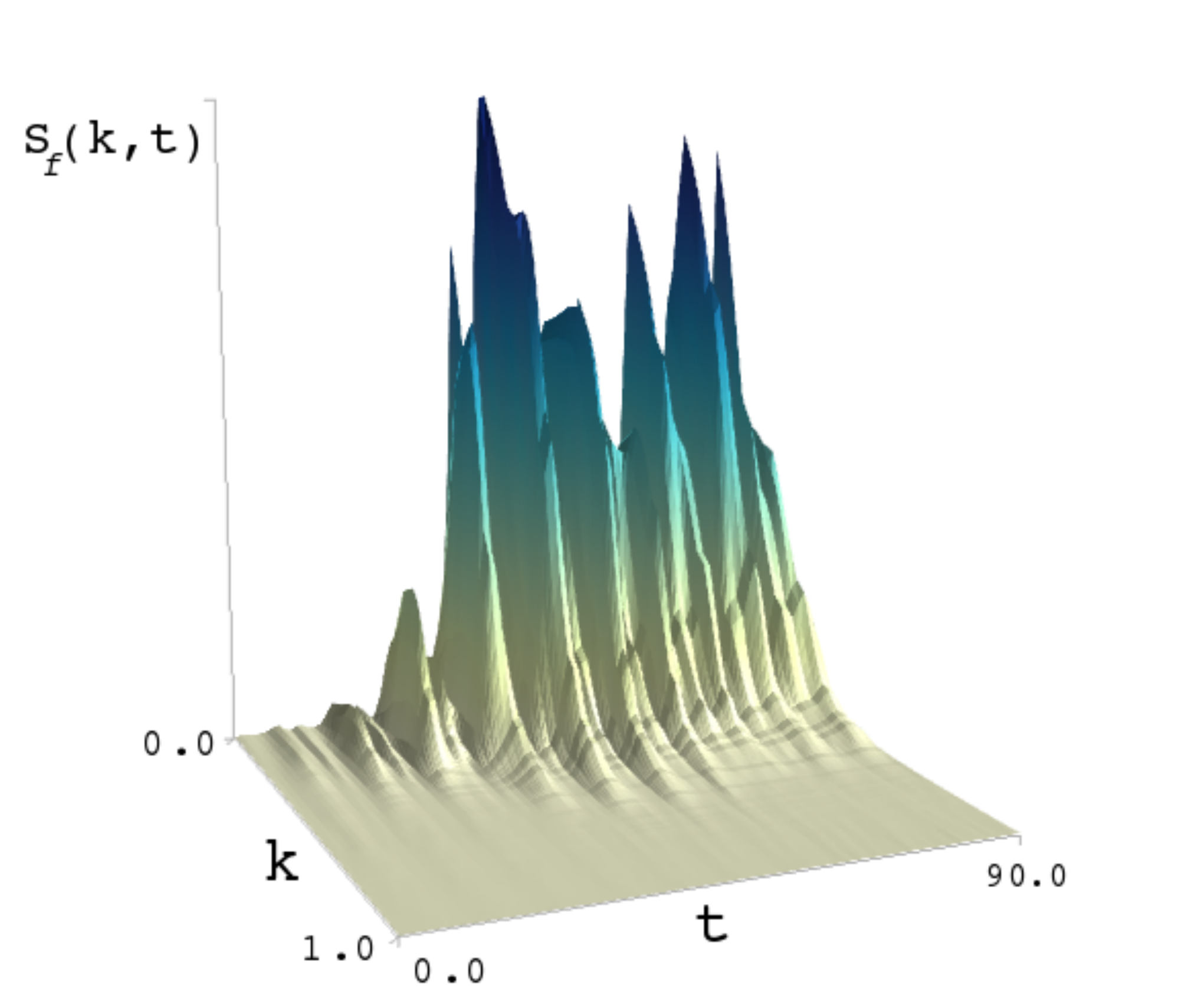}
\includegraphics[width=0.49\linewidth]{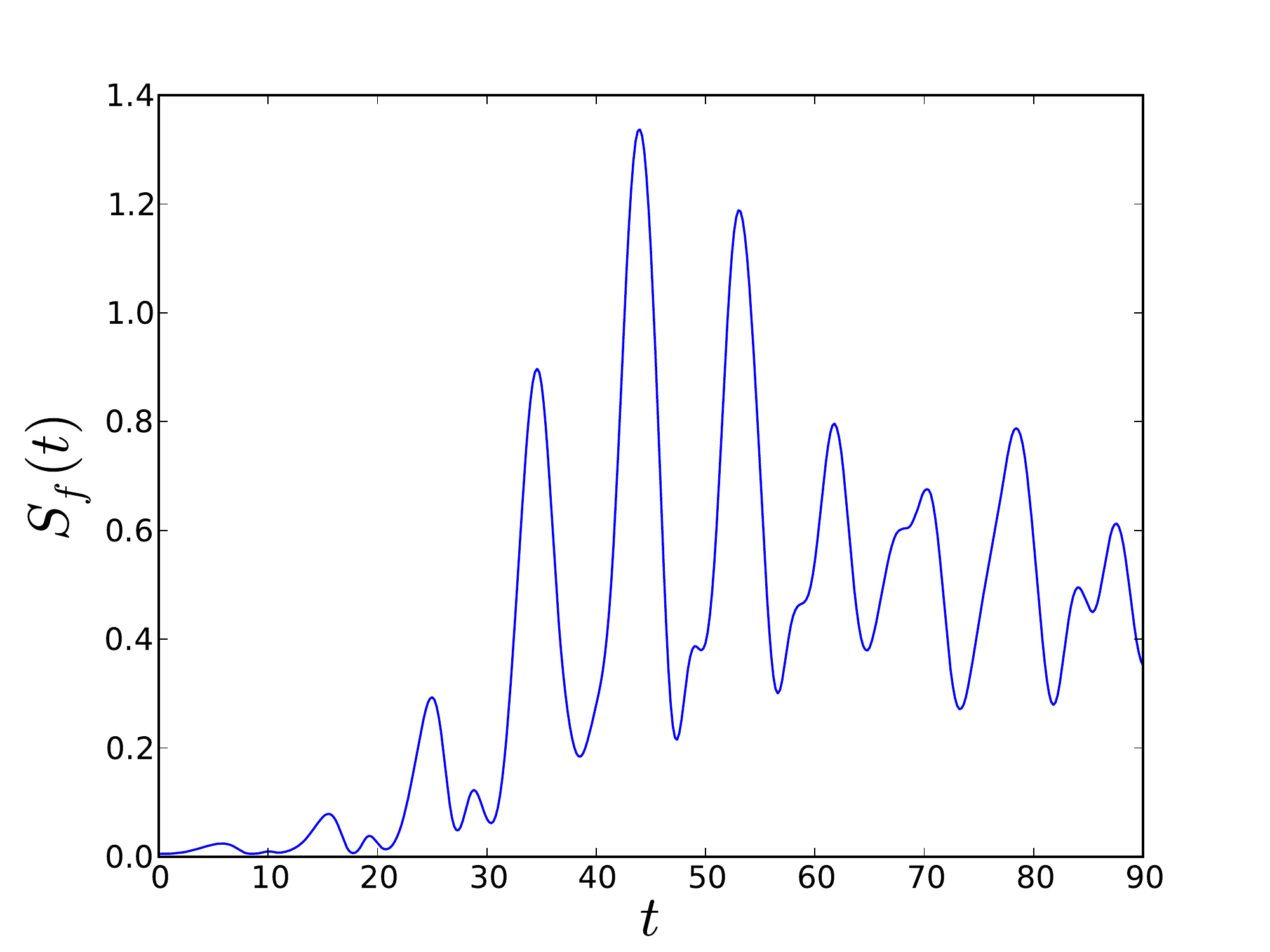}
\caption{Relative entropy density $S_f(|k|,t)$ and relative entropy $S_f(t)$ as a function of time for a simulation of initial temperature $T=0.25$. Initially the relative entropy is zero, corresponding to the time just after the quench when the system is still in equilibrium. As parametric resonance takes place, spikes in the low $k$ part of the relative entropy density emerge, signaling the formation of oscillons. The amplitude of the spikes correlate with the number of oscillons formed, as explained below. As the system evolves back to equilibrium, the oscillons disappear and the relative configurational entropy goes back to zero. The wave vector magnitude $k$ has units of $m$ and time has units of $m^{-1}$.}
\label{Fig:entropyInTime}
\end{figure}

In Fig.~\ref{Fig:entropyInTime} we plot the RCE density (left) and RCE (right) as a function of time for the same simulation with $T=0.25$ of Figs.~\ref{Fig:spectrum} and \ref{Fig:oscillons}. Since the field starts in thermal equilibrium, the RCE density, $S_f(|k|,t)$, is zero everywhere initially. After the quench, energy is transferred to low-$k$ modes and spikes in $S_f(|k|,t)$ begin to appear. These are clearly seen after integrating over the $k$ modes, as shown in the plot of the RCE on the right. 

\begin{figure}[htbp]
\includegraphics[scale=0.5]{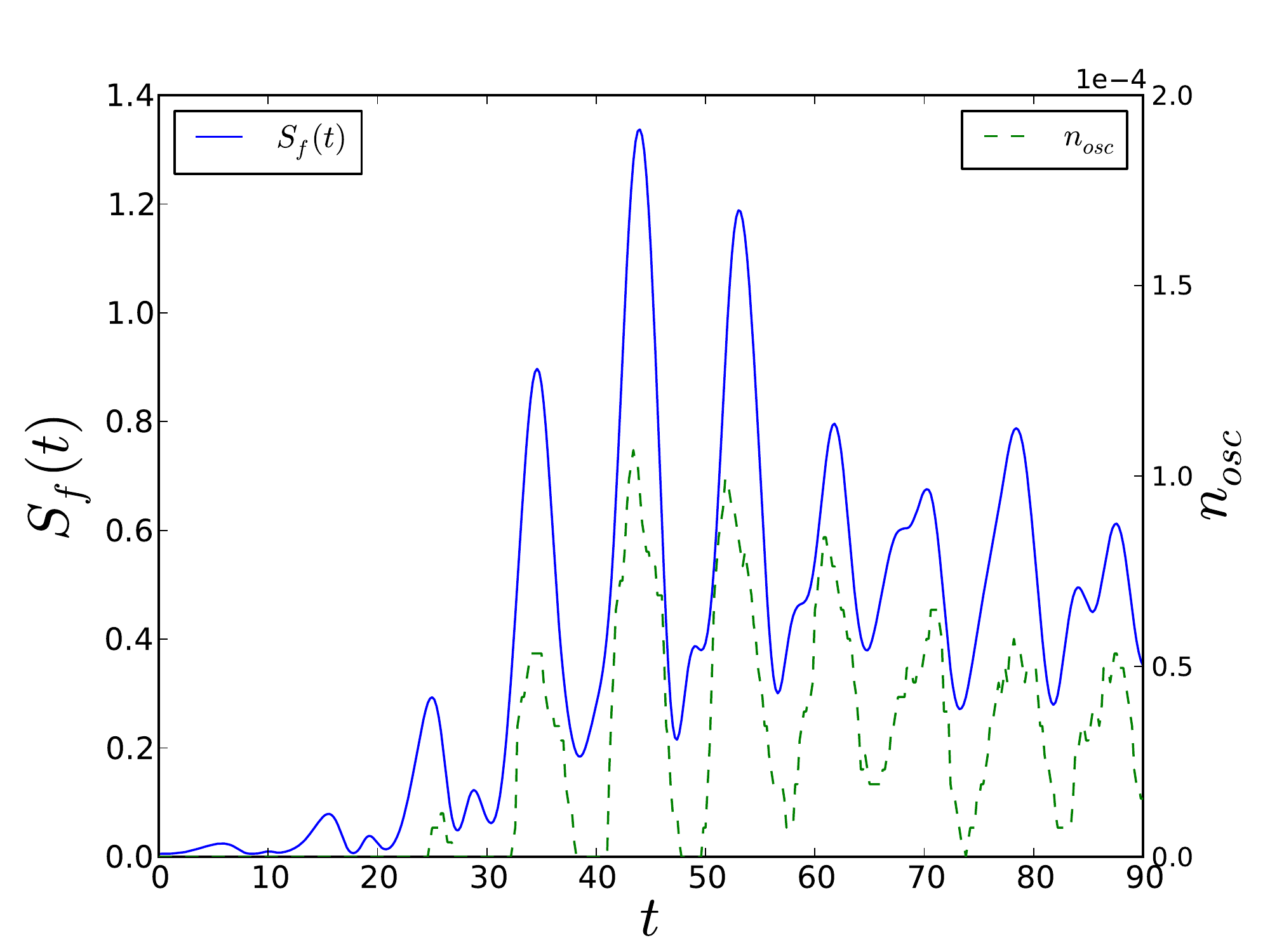}
\caption{Relative configurational entropy $S_f(t)$ (continuous line) and number density of oscillons $n_{\rm osc}(t)$ (dashed line) as a function of time for a simulation of initial temperature $T=0.25$. Initially, both the relative entropy and the number density of oscillons are zero, since the system starts in equilibrium. After the quench, a clear correlation is seen between the spikes in $S_f(t)$ and the maxima of $n_{\rm osc}$. The wave vector magnitude $|k|$ has units of $m$ and time has units of $m^{-1}$.}
\label{Fig:entropy_oscillons}
\end{figure}

In order to substantiate the claim that the RCE gives a quantitative measure of the emergence of spatiotemporal structure in the system, in Fig.~\ref{Fig:entropy_oscillons} we plot both $S_f(t)$ and the number density of oscillons $n_{\rm osc}(t)=N_{\rm osc}(t)/V$. It is quite clear that the spikes in $S_f(t)$ coincide with the synchronous emergence of oscillons, and that the higher the amplitude of $S_f(t)$ the larger the number density of oscillons present. (An advantage of defining the relative entropy to be dimensionless is that it can be consistently used for similar simulations in lattices of different size. When the simulation size is increased while working with the same initial temperature, the power spectrum will look the same but the number of oscillons formed will be proportionally larger. Therefore, our definition of the relative entropy is a consistent measure of the number density of oscillons formed.)

As an illustration, the highest peak in the two-point correlation function, shown in the right-side plot of Fig.~\ref{Fig:spectrum}, appears at time $t=44m^{-1}$.  The isosurface snapshot of Fig.~\ref{Fig:oscillons} shows the richness of spatial structure in the field at that time. This is also the time when the highest peak in the RCE appears, and corresponds to the maximum in the number density of oscillons, as is clear from Fig.~\ref{Fig:entropy_oscillons}. Eventually, as the system evolves towards its final equilibrium state, the relative entropy goes slowly back to zero, signaling the disappearance of coherent structures in the field. 

For low temperatures, the Hartree approximation is valid for all times and Eq.~\ref{equilibrium-lattice} describes well the equilibrium spectrum. For larger temperatures ($T\gtrsim 0.30$), the Hartree approximation breaks down and we cannot use Eq.~\ref{equilibrium-lattice} as descriptive of the equilibrium state of the system. Although it will still be a good base for studying the emergence of structure in the field, it should be used with care.  

\begin{figure}[htbp]
\includegraphics[scale=0.5]{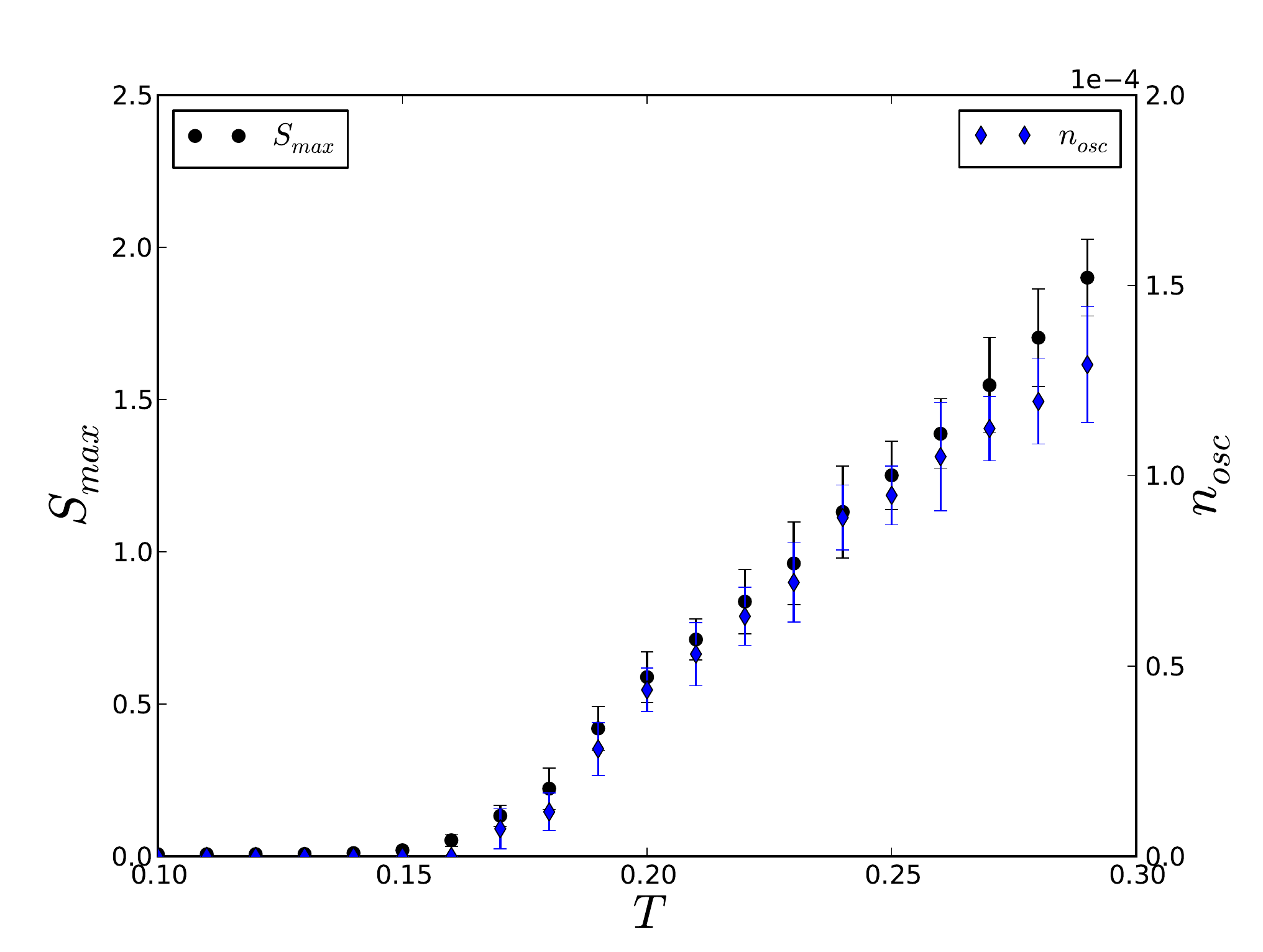}
\caption{Ensemble-averaged values of the maxima of the RCE and of the oscillon number density for a range of different initial temperatures. For low temperatures $T\lesssim 0.15$, no oscillon formation is possible and the RCE is zero. As the temperature increases, the maxima of the RCE increase in amplitude at the same rate as the maxima of the oscillon number density. For temperatures $T\sim0.28-0.29$ the rates begin to diverge as the Hartree approximation is no longer valid.}
\label{Fig:entropy_number_density}
\end{figure}

To verify that the correspondence between the RCE and the number density of oscillons holds for the temperature range where oscillons appear in the system, we extract the maxima of the RCE and the corresponding oscillon number density at that time for a range of initial temperatures $0.10\leq T\leq 0.29$. (For example, for $T=0.25$ this would be at $t=44m^{-1}$ in the simulation displayed in Fig.~\ref{Fig:entropy_oscillons}.) For each value of $T$ we perform an ensemble average over $15$ simulations and plot the results in Fig.~\ref{Fig:entropy_number_density}. The vertical axis on the left displays the ensemble-averaged maximum value of the RCE, $S_{\rm max}$, with the data shown in circles, while the right vertical axis displays the corresponding ensemble-averaged oscillon number density $n_{\rm osc}$ with the data depicted by diamonds. The error bars show the standard deviation of the ensemble. For low temperatures $0.10\leq T\leq 0.15$, the fluctuations on the field do not have large enough amplitude to lead to oscillon formation and the RCE is zero. As the temperature increases, parametric amplification of the oscillon-related $k$ modes triggers the formation of oscillons. The relationship between the RCE and the number density of oscillons is evident: higher temperature leads to both higher RCE and larger oscillon number-density. For temperatures $T\sim0.28-0.29$ we begin to see a discrepancy between the two values, the reason being that the Hartree approximation starts to diverge from the true equilibrium spectrum. (For reference, symmetry restoration occurs roughly at $T\sim 0.32$.) We note that the results of Fig.~\ref{Fig:entropy_number_density} are independent of the simulation volume.

\section{Conclusion and Discussions}

We investigated the nonequilibrium dynamics of symmetry breaking in the context of a $3d$ real scalar field model with a double-well potential. Preparing the system in a parity-invariant initial thermal state, we break the ${\cal Z}_2$ symmetry by adding a cubic term to the potential. As a result, coherent spatiotemporal structures emerge, an ensemble of long-lived oscillons. We proposed a measure to quantify the emergence of spatiotemporal order, which essentially counts the modes out of equilibrium. This measure, which we called relative configurational entropy following our previous work of Ref. \cite{GleiserStamEnt}, provides an accurate description both of the departure from equilibrium and of the emergence of coherent structures in the field. ``Relative'' here refers to computing an entropic distance in field configuration space from a baseline which we took to be the thermal state. This way, we are able to provide the informational content of nonequilibrium field structures, in particular of coherent states that emerge during spontaneous symmetry breaking. We could just as easily have studied the information content of domain-wall formation had we used a different initial state centered at the maximum of the double-well potential.

The measure we proposed here should be generalizable to many different contexts. For example, it should be possible to apply it to models with gauge fields, or in nonrelativistic applications of interest to condensed matter physics, such as Ginzburg-Landau models of superfluids and superconductors. Furthermore, a similar measure should also be of interest in cosmological contexts where topological and nontopological structures appear due to symmetry breaking driven by the cosmic expansion. Finally, it would be interesting to see if such ideas could be extended to compute the relative configurational entropy of metric spaces, thus providing a possible measure of gravitational entropy. Work along these lines is currently in progress.

\acknowledgments
MG is supported in part by a National Science Foundation grant PHY-1068027. NS is a Gordon F. Hull Fellow at Dartmouth College.

\section{Appendix: Analytical Estimate of the Modal Fraction Normalization at Equilibrium}
The normalization of the modal fraction at equilibrium in Eq.~\ref{equilibrium_modal} can be computed analytically. This will provide an estimate of how important lattice effects are for different lattice parameters. We assume spherical symmetry in momentum space and impose a UV cutoff $\Lambda$ to perform the integral

\begin{equation}
 \int |\phi_{\rm eq}^{\rm an}(k)|^2 d^3k=\int_0^{\Lambda}\frac{T}{k^2+m_H^2}4\pi k^2dk=4\pi T\left(\Lambda-m_H\tan^{-1}\frac{\Lambda}{m_H}\right),
\label{analytical_approx}
\end{equation}
where $T$ is the temperature and $m_H$ is the Hartree mass. The value of the UV cutoff $\Lambda$ is chosen appropriately to describe the discretization employed in the numerical calculations. Here, we pick the value $\Lambda=\sqrt{3}\pi/dx$ which corresponds to the largest wavevector represented in a cubic lattice of lattice spacing $dx$. Geometrically, this choice makes the integration volume of Eq.~\ref{analytical_approx} to be the smallest sphere that fully covers the cubic lattice in momentum space.

We evaluate Eq.~\ref{analytical_approx} for $T=0.25$ and several choices of lattice spacing $dx$. The results are shown in Table~\ref{approximation_results}. For large values of $dx$, the analytical approximation is in good agreement with the numerically calculation; but as the lattice spacing is decreased, the two values begin to diverge. To see why the analytical approximation underestimates the value of the integral, note that $k^2$ grows faster than the numerically-calculated $k_{\textrm{eff}}^2$ of Eq.~\ref{k_eff} and this becomes increasingly more noticeable for smaller lattice spacings and hence larger volume of integration. This further justifies our incorporation of lattice correction effects in the dispersion relation in Eq.~\ref{equilibrium-lattice}, as the discrepancy is of the order of $10\%$ for $dx=0.5$ used throughout this paper.

\begin{table}[htdp]
\begin{center}
\begin{tabular*}{0.5\textwidth}{@{\extracolsep{\fill}} |c||c|c|} \hline
$dx$   &$\int |\phi_{\rm eq}^{\rm latt}(k)|^2 d^3k$ &  $\int |\phi_{\rm eq}^{\rm an}(k)|^2 d^3k$  \\ \hline
0.9    & 14.8874   & 14.3853       \\ \hline
0.7    & 20.9985   & 19.6331       \\ \hline
0.5    & 32.0204   & 29.1701       \\ \hline
0.3    & 57.611    & 51.5837       \\ \hline
0.1    & 184.715   & 164.262       \\ \hline
\end{tabular*}
\end{center}
\caption{Comparison between the numerical and analytical values of the modal fraction normalization at equilibrium for different values of lattice spacing. The analytical approximation given by Eq.~\ref{analytical_approx} works well as long as the lattice spacing is not very small. For smaller values of $dx$, the lattice effects on the dispersion relation calculated in Eq.~\ref{k_eff} have to be taken into account. The lattice spacing $dx$ is given in units of $m^{-1}$ and the normalization factors have units of $[Tm]$.}
\label{approximation_results}
\end{table}%

\vfill\eject


\begin{thebibliography}{99}

\bibitem{Scott} S. Scott, {\it Nonlinear Science:Emergence and Dynamics of Coherent Structures} (Oxford University Press, Oxford, UK, 2003).

\bibitem{Rajamaran} R.~Rajamaran, {\it Solitons and Instantons} (North-Holland, 1987).

\bibitem{Nelson} D. R. Nelson, {\it Defects and Geometry in Condensed Matter Physics} (Cambridge University Press, Cambridge, UK, 2005).


\bibitem{Peskin} M.E. Peskin and D.V. Schroeder, {\it An Introduction to Quantum Field Theory} (Addison-Wesley, Reading, MA, 1995).

\bibitem{Vilenkin} T.~W.~B.~Kibble, J.\ Phys.\ A {\bf 9}, 1387 (1976); A.~Vilenkin and E.~P.~S. Shellard,
{\it Cosmic Strings and Other Topological Defects} (Cambridge
  University Press, 1994).
  
\bibitem{Lee} T.D. Lee and Y. Pang, Phys. REp. {\bf 221}, 251 (1992).

\bibitem{Coleman}S. Coleman, Nucl.\ Phys.\ B {\bf 262}, 263 (1985);
  [Erratum-ibid.\ B {\bf 269}, 744 (1986)].
  
\bibitem{Friedberg} R.~Friedberg, T.~D. Lee, and A.~Sirlin, Phys. Rev. {\bf D13}, 2739 (1976).

\bibitem{Bogolyubsky} I.~L. Bogolyubsky and V.~G. Makhankov, JETP Lett. {\bf 24}, 12 (1976).

\bibitem{Gleiser1} M.~Gleiser, Phys. Rev. {\bf D49}, 2978 (1994); E.~J. Copeland, M.~Gleiser, and H.~R. Muller, Phys. Rev. {\bf D52}, 1920 (1995), hep-ph/9503217.

\bibitem{Oscillons} M.~Hindmarsh and P.~Salmi, Phys. Rev. {\bf D74}, 105005 (2006), G.~Fodor, P.~Forgacs, P.~Grandclement, and I.~R\'acz, Phys. Rev. {\bf D74}, 124003 (2006); G. Fodor, P. Forg\'acs, Z. Horv\'ath, and \'A. Luk\'acs, Phys. Rev. D {\bf 78}, 025003 (2008); G. Fodor, P. Forg\'acs, Z.
Horv\'ath, and M. Mezei, Phys. Rev. D {\bf 79}, 065002 (2009); {\it ibid.}, Phys.
Lett. B {\bf 674}, 319 (2009).

\bibitem{Gleiser2} M. Gleiser, Phys.\ Lett.\ B {\bf 600}, 126 (2004); P.~M. Saffin and A.~Tranberg, J. of High Energy Physics {\bf 2007}, 030 (2007).

\bibitem{GleiserThor} M.~Gleiser and J.~Thorarinson, Phys. Rev. D   {\bf 76}, 041701 (2007).

\bibitem{GrahamMIT} E.~Farhi, N.~Graham, V.~Khemani, R.~Markov and R.~Rosales, Phys.\ Rev.\ D {\bf 72}, 101701 (2005).

\bibitem{GleiserThor2} M. Gleiser, B. Rogers, and J. Thorarinson, Phys. Rev. D{\bf 77}, 023513 (2008).

\bibitem{GrahamSM} N.~Graham, Phys. Rev. Lett. {\bf 98}, 101801 (2007); [Erratum-ibid. {\bf 98}, 189904 (2007)].

\bibitem{GleiserHowell} M. Gleiser and R. Howell, Phys. Rev. E {\bf 68}, 065203(RC) (2003); M. Gleiser and R. Howell, Phys. Rev. Lett. {\bf 94}, 151601 (2005).

\bibitem{GGS1} M. Gleiser, N. Graham, and N. Stamatopoulos, Phys. Rev. D {\bf 82}, 043517 (2010).

\bibitem{GGS2} M. Gleiser, N. Graham, and N. Stamatopoulos, Phys. Rev. D {\bf 83}, 096010 (2011).

\bibitem{Amin} M. A. Amin, R. Easther and H. Finkel, JCAP, 1012:001 (2010); M. A. Amin and D. Shirokoff, Phys. Rev. D {\bf 81} 085045 (2010); M. A. Amin, [arXiv:1006.3075]; M.A. Amin, R. Easther, H. Finkel, R. Flauger, and M. P. Hertzberg, [arXiv:1106.3335].

\bibitem{GleiserSic} M. Gleiser and D. Sicilia, Phys. Rev. Lett. {\bf 101}, 011602 (2008); {\it ibid.} Phys. Rev. D {\bf 80}, 125037 (2009).

\bibitem{Shannon} C. E. Shannon, The Bell System Technical J. {\bf 27}, 379 (1948); {\it ibid.} 623 (1948).

\bibitem{Brillouin} L. Brillouin, {\it Science and Information Theory} (Academic Press, New York, 1956).

\bibitem{GleiserStamEnt} Marcelo Gleiser, Nikitas Stamatopoulos, [arXiv:1111.5597].

\bibitem{langevin} M. Morikawa, Phys.Rev. D {\bf 33}, 3607 (1986).

\bibitem{isotropic} M. Patra and M. Karttunen, Num. Meth. for PDEs {\bf 22}, 936, (2005).

\bibitem{hartree} G. Aarts, G. F. Bonini, and C. Wetterich, Phys. Rev. D{\bf 63}, 025012 (2000); G. Aarts, G. F. Bonini, and C. Wetterich, Nucl. Phys. B {\bf 587}, 403 (2000).

\bibitem{lattice-dispersion} Nikitas Stamatopoulos, [arXiv:1201.3368].

\bibitem{simulation} http://www.youtube.com/watch?v=wEjZT-fg7EY

\bibitem{kullback-leibler} S. Kullback and R. A. Leibler, Ann. Math. Stat. {\bf 22}, 79, (1951).

\end{thebibliography}
\end{document}